*Frequency of Coronal Mass Ejection Impacts with Early Terrestrial Planets and Exoplanets Around Active Solar-like Stars*

Christina Kay[1,2], Vladimir S. Airapetian[1,3], Theresa Lüftinger[4], Oleg Kochukhov[5]

[1]*NASA GSFC/SEEC*, [2]*Catholic University of America*, [3]*American University, DC*, [4]*University of Vienna, Austria*, [5]*Uppsala University, Sweden*

**Abstract**

Energetic flares and associated coronal mass ejections (CMEs) from young magnetically active solar-like stars can play a critical role in setting conditions for atmospheric escape as well as penetration of accelerated particles into their atmospheres that promotes formation of biologically relevant molecules. We have used the observationally reconstructed magnetic field of the 0.7 Gyr young Sun's twin, *k¹ Ceti*, to study the effects of CME deflections in the magnetic corona of the young Sun and their effects on the impact frequency on the early Venus, Earth, and Mars. We find that the coronal magnetic field deflects the CMEs toward the astrospheric current sheet. This effect suggests that CMEs tend to propagate within a small cone about the ecliptic plane increasing the impact frequency of CMEs with planetary magnetospheres near this plane to ~ 30% or by a factor of 6 as compared to previous estimate by Airapetian et al. (2016). Our model has important implications for the rise of prebiotic chemistry on early terrestrial planets as well as terrestrial-type exoplanets around young G-K dwarfs.

1. Introduction

The *Hubble Space Telescope (HST)*, *Kepler Space Telescope*, and recent *Transiting Exoplanet Survey Satellite (TESS)* observations suggest that young and active F-M-type planet hosts are magnetically active stars showing evidence of strong surface magnetic fields, large starspots, and intense X-ray coronal emission. Many of them also show signatures of frequent superflares, explosive phenomena in the stellar atmosphere associated with an abrupt release of magnetic energy 10-1000 times larger than that of the largest solar flare (class X28 +) ever observed on the Sun (Schaefer et al. 2000; Maehara et al. 2012; 2015). Energetic solar flares (*GOES* X class) with energies E > $3.5 \times 10^{29}$ erg are accompanied by coronal mass ejections (CMEs; Yashiro and Gopalswamy 2009). This may suggest that flares from young Sun analogs should also be associated with energetic stellar CMEs, which could have been an important factor for planetary habitability in the early history of our solar system and/or most extrasolar systems (Maehara et al. 2012; Shibayama et al. 2013; Takahashi et al. 2016; Notsu et al. 2019; Airapetian et al. 2016; 2019a). Drake et al. (2013) applied the empirical solar flare-CME scaling to young solar-type stars and concluded this correlation will output unrealistically high CME-driven mass loss. The possible resolution of this problem may come from the recent magnetohydrodynamnic (MHD) simulations of stellar compact CMEs initiated from active regions Alvarado-Gomez et al. (2018). The authors concluded that CMEs with energy less than 3 x $10^{32}$ erg (the Carrington scale CME event) should be confined by a large-scale surface magnetic field of 75 G that is typical for these stars (Airapetian et al. 2019b).

Recent MHD simulations of the initiation and launching of CMEs into the stellar corona of a young solar twin, *k¹ Ceti*, suggest that shearing flows can also energize the stellar streamer-belt followed by the eruption of a flux rope, releasing ~7×10³³ erg of magnetic free energy in ~10 hours (Lynch et al. 2019). Magnetic reconnection during the stellar flare creates the twisted flux rope structure of the ejecta and the ~2000 km/s eruption drives a strongly magnetized shock.

The powerful CME events from the current Sun are the most geoeffective space weather events in the Solar system impacting magnetospheres, ionospheres and thermospheres of Earth and solar system bodies. For terrestrial-type exoplanets orbiting the young Sun and active stars, the dynamic and magnetic pressure from powerful (Carrington-type) CMEs can push the day-side planetary magnetosphere to a stand-off distance of the less than two planetary radii (Khodachenko et al. 2007; See et al. 2014; Airapetian et al. 2016; Kay et al. 2016; Garraffo et al. 2017; Patsourakos and Georgoulis 2017). This widens the polar cap by opening 70% of the magnetospheric field and facilitating entry for energetic particles accelerated by CME-driven shocks. These energetic particles penetrate the planetary atmosphere igniting ionization driven by secondary electrons. The secondary electrons at energies of ~ 10 eV are instrumental in dissociation of molecular nitrogen, carbon dioxide, methane and water vapor and thus forming free radicals including NO, $NO_2$, CO, CH, CN and other molecules that form abundant nitrous oxide, the potent greenhouse gas, and hydrogen cyanide, the feedstock molecule of life, in the upper troposphere-stratosphere of the planet (Airapetian et al. 2016; 2019a). We have previously shown that the combination of Extreme UV emission from quiescent hot stellar coronae and frequent, powerful stellar flares may have significant effect on atmospheric escape, and thus is detrimental for habitability of close-in exoplanets around M dwarfs over the course of a few tens of Myrs (Airapetian et al. 2017; Garcia-Sage et al. 2017; Dong et al. 2018). However, for early terrestrial planets and other Earth-sized exoplanets around active G-K stars at Venus and Earth distances, frequent CME collisions may provide fertile ground for biogenic conditions on. Thus, accurate characterization of the frequency of CME impacts with exoplanetary magnetospheres is an important task.

Young (300-700 Myr) solar-type (G-type) stars are known to generate powerful superflares with energies > 2 x 10³⁴ erg up to 5 x 10³⁵ erg with the frequency of superflares following a power-law distribution $N(E>E_0) \sim E^{-1}$ (Notsu et al. 2019). This is consistent with the detection of a 2 x 10³⁴ erg superflare observed on the young Sun analog, a 750 Myr G5V, κ¹Cet (Robinson & Bopp 1987; Schaefer et al. 2000). The Kepler flare data complemented with Gaia data suggest that the young solar analogs with ages of 300-600 Myr generate supeflares at energies > 2 x 10³⁴ erg.

Assuming that CMEs from the young Sun at 0.5-0.7 Gyr are launched isotropically into the interplanetary space, Airapetian et al. (2016) estimated the frequency of impacts with the early Earth as ~ 1 event per day, which suggests an efficient supply of organic molecules via CME-driven energetic particles. However, this estimate does not account for deviations of CME trajectories from the purely radial trajectory revealed by the change in their latitude and longitude (deflections) observed in the solar corona. These deflections from the radial path are driven by magnetic forces, which direct the CME motion away from coronal holes toward the astrospheric current sheet (ACS), the region where solar coronal magnetic field reverses its radial polarity (Gopalswamy et al. 2009; Kay et al. 2015, 2016). A recent study of the impact of CMEs from an M4 dwarf suggested that deflection toward the ACS can be important in increasing the frequency of impacts by a factor of 10 (Kay et al. 2016), suggesting that the frequency found by Airapetian et al. (2016) may be an underestimate.

In this Letter, we use the spectropolarimetric observations of one of the best proxies for the young Sun at the time when life started on Earth (Do Nascimento et al. 2016; Airapetian et al. 2019b) and apply our CME deflection model, Forecasting a CME's Altered Trajectory (*ForeCAT*) model to characterize the coronal magnetic environment and the role of CME deflection on the likelihood of CME collisions with the early Earth and other Earth-sized exoplanets around G-K dwarfs. Section 2 provides the model description and its setup condition. Section 3 describes the *ForeCAT* model results for $k^1$ *Ceti*. In Section 4 we present the CME probabilistic model developed from the *ForeCAT* model results. Section 5 discusses the implications of the model results for the habitability of early terrestrial planets and Earth-sized exoplanets around active stars.

## 2. Model Setup

To model the CME trajectories in the stellar corona and in the inner astrosphere, we used a three-dimensional (3D) *ForeCAT* model (Kay et al. 2013, 2015) that simulates the behavior of a rigid torus, intended to represent the flux rope of a CME, as a function of radial distance. We specify the CME's initial position on the stellar disk (latitude and longitude), the tilt measured clockwise with respect to the equatorial plane, the shape, and size of CME. We provide ForeCAT with an empirical model for the CME mass, angular width, and velocity as a function of radial distance. Typically for all three parameters we assume a profile that rapidly increases in the low corona then remains constant beyond about 5 stellar radii, which is consistent with solar observations (e.g. Patsourakos et al. 2010a, 2010b; Aschwanden 2009). The simulated CME is embedded in a magnetic background that fully determines its deflection and rotation from magnetic forces, causing changes in the latitude, longitude, and orientation of the torus as it propagates outward.

Our recent theoretical study of the young Sun's twin, $k^1$ *Ceti*, suggests that over the course of 11 months its global corona underwent a drastic transition from nearly a dipole to a complex magnetic topology strong dipole quadrupole and octopole components and the formation of a low-latitude coronal hole. We used the stellar magnetograms reconstructed for two epochs, 2012.9 and 2013.8, using high-resolution spectropolarimetric data in Stokes I and V obtained with dedicated instrumentation including the NARVAL spectropolarimeter at the Telescope Bernard Lyot (France) and HARPSpol@ESO (Chile) (*Rosén et al. 2016)*. We use the low-resolution reconstructed large-scale magnetic field ($l_{max}$ = 10) of the star to determine the harmonic coefficients and construct a Potential Field Source Surface model of the background magnetic field at the source-surface height of 2.2 $R_{sun}$ (Rosén et al. 2016). As discussed in Airapetian et al. (2019b), the global magnetic field of $k^1$ *Ceti* at 2012.9 is mostly dipolar with $9^0$ tilt with respect to the rotation axis and resembles the current Sun's global field at solar minimum. However, the magnetic field shows the signatures of large-scale restructuring to $45^0$ tilted dipolar component with 2/3 of the contribution from quadrupolar and octopolar components, which is typical for the declining of the declining phase of solar maximum. To determine the background solar wind density, we use the Guhathakurta et al. (2006) model that empirically scales the value based on the distance from the ACS and the radial distance. By assuming constant mass flux, we can then calculate the radial solar wind speed. In our empirical models for the background solar wind we use the same coefficients as previously used for solar ForeCAT simulations.

We expect the CMEs from *k¹ Ceti* to differ slightly from those from the current Sun. Here we consider a range of CMEs between $10^{16}$ and $10^{17}$ g, similar to but toward the more massive side of typical solar CMEs. Using empirical relations derived from 45 well-constrained CMEs studied in Kay & Gopalswamy (2017) we relate the speeds, $V_{CME}$ (in km/s), and angular widths, AW (in °), to the CME masses, $M_{CME}$ (in g).

$$V_{CME} = 660 \log_{10}(M_{CME}) - 9475$$

$$AW = 39.6 \log_{10}(M_{CME}) - 540$$

The corresponding CME speeds vary between 1070 km/s and 1730 km/s and the angular widths vary between $96^0$ and $126^0$. Note that the angular width is the full angular width, not the half-angular width often reported from solar reconstructions.

### 3. ForeCAT Simulations

We performed *ForeCAT* simulations for both epochs of *κ¹ Ceti* magnetic field observations at 2012.9 and 2013.8. For each epoch, we select seven different initial CME locations. Each initial location corresponds to a polarity inversion line in the background magnetic field at the stellar surface. From each initial location we simulated six CMEs spanning the range of CME masses between $10^{16}$ and $10^{17}$ g. The upper and lower panels of Figure 1 show the results of the *ForeCAT* simulations for the 2012.9 epoch and the 2013.8 epoch, respectively, illustrating where impacts are most likely to occur. The faint red-blue color contours in the background show the radial magnetic field at the stellar surface with white indicating values near zero. We initiate CMEs along polarity inversion lines, locations where the surface magnetic field reverses direction, so along the white lines in the background contours. The colored stars indicate the center of the seven different initial locations in each epoch. Each initial position is shown with a different color. The precise color has no meaning; rather it is only a means to differentiate between origins. The line contours show the total magnetic field strength at the source-surface height with the darkest contours indicating the weakest values and the location of the ACS. We expect the CMEs to deflect away from their initial positions and toward the ACS. The circles represent the final location of the center of the deflected CMEs are colored the same as the star representing their initial position. The circle size corresponds to the CME mass with larger circles being more massive CMEs. To determine the likelihood of impact we need the full extent of the CME, not just the position of the center. For each simulation, we show the projection of the CME's final position and orientation onto the stellar surface. Each CME has a torus shape of uniform cross-sectional width but this projects to nonuniform shapes on the flattened spherical surface, particularly near high latitudes. The yellow-purple color contours show the overlap of the projections for all simulations and are scaled such that if one CME were to erupt from each star the value indicates the expected number of impacts at that location. There are locations where the number of expected impacts exceeds one because several initial positions produce CMEs that reach that location and the maximum number would be seven as we consider seven initial locations.

The global stellar magnetic field at 2012.9 epoch is mostly dipole-like with a $9^0$ tilt with respect to the ecliptic plane, and thus resembles a solar minimum configuration with a relatively flat ACS. Many of the initial locations are directly underneath the latitudes and longitudes where the ACS appears at higher radial distances (e.g. the dark and light blue, green, and orange cases).

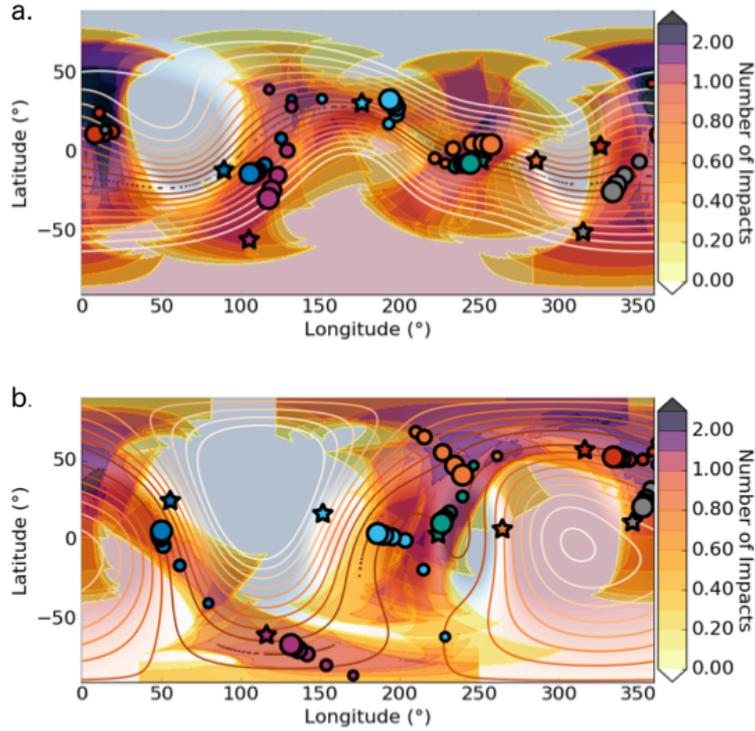

*Figure 1: Final positions and orientations of CMEs from κ¹ Ceti and the expected number of impacts for epoch 2012.9 (top) and 2013.8 (bottom). The faint red-blue contours show the surface magnetic field and the line contours show the magnetic field farther out. Each CME ensemble is represented with a different color with the star indicating the initial location and the circles representing the final position with circle size increasing with CME mass. The yellow-purple color contours show the overlap of the projections for all simulations and are scaled such that if one CME were to erupt from each star the value indicates the expected number of impacts at that location.*

This may result from the low resolution of our magnetogram ($l_{max}$=10). For the most part, we cannot resolve the structure of individual active regions, so the polarity inversion lines at the surface are more representative of the global structure. In this case, these CMEs are similar to streamer blowout CMEs observed on the Sun or the simulation of Lynch et al. (2019). For these cases, we see some deflection but the CMEs tend to remain near the ACS. The initial locations of the gray and purple CMEs are slightly farther away from the ACS. For these cases we can see a clear deflection of the CMEs toward the ACS. As seen for solar CMEs, the deflection tends to decrease with increasing CME mass. Unlike M dwarfs with stronger magnetic backgrounds, the CMEs are not trapped at the ACS. The least massive CMEs deflect fully to the ACS, but then can continue moving beyond it as the magnetic gradients are not sufficiently strong to stop their motion. The purple $10^{16}$ g CME (smallest circle) deflects almost $85^0$ northward below $1.5 R_\odot$, placing it near the ACS at a longitude of $135^0$. The rate of deflection begins decreasing and the CME reaches a peak latitude of $43.3^0$ at a distance of $2.3 R_\odot$, after which the motion reverses and slowly decreases to a final latitude $39.0^0$ at $50 R_\odot$. The magnetic gradients clearly oppose the continued deflection beyond the ACS, slowing it down and eventually reversing it, but are insufficient to counter the CME's momentum and trap it at the ACS. The CMEs denoted as red are initiated at the only bipolar active-region-like structure in epoch 2012, showing the effects of the smaller scale structure. The western flux system (right side using solar conventions) is weaker than the eastern one (left side) causing these CMEs to exhibit a westward deflection (rightward across $360/0^0$ boundary in Figure 1). The CMEs continue deflecting until they reach a region of strong magnetic field around $25^0$ longitude, which halts the deflection of even the most-massive CMEs. Deflection toward the ACS minimizes the net forces acting upon a CME, so similarly we would expect rotation toward the ACS's orientation to minimize the net torque upon a CME. For epoch 2012.9, however, we do not see a consistent trend in the rotation. Some cases rotate toward the orientation of the nearby portion of the ACS, while others do not.

For epoch 2013.8, the magnetic field has a complex orientation resembling a dipole field tilted at $45^0$ with significant (~50%) contribution from multipole components. This topology is characteristic for the current Sun's global magnetic field at solar maximum. The ACS is correspondingly inclined, extending up to high latitudes. The surface field shows a few smaller bipolar regions, but we still cannot resolve individual active regions and the majority of the initial locations are again directly underneath the ACS. We use the same set of colors for the 2013.8 epoch as the 2012.9 epoch, but there is no connection between CMEs of the same color in different epochs. The blue, purple, green, and red cases all begin underneath the ACS and any deflection of these individual CMEs simply moves them along the ACS. The light blue, orange, and gray cases begin slightly farther from the ACS and deflect toward it with some of the least massive CMEs being able to push slightly beyond it.

We find that for 2013.8 epoch the final orientations tend to be aligned with the local orientation of the ACS. The impact contours trace out the orientation of the ACS for all longitudes, unlike epoch 2012.9 where the CMEs are nearly perpendicular at some longitudes. As for the deflection, the rotation tends to decrease with increasing CME mass. For each mass, the average rotation in epoch 2013.8 is slightly smaller that of epoch 2012.9, but not terribly dissimilar. The final alignment with the ACS is likely due the relative orientation of the ACS and the initial polarity inversion lines. The difference between the two tends to be much smaller for epoch 2013.8 due to the higher inclination of the ACS.

To develop a description of the average deflection of *κ1 Ceti* CMEs we study for a relation between the CME mass and the final distance from the ACS. Figure 2 shows these results for epoch 2012.9 (top panel) and epoch 2013.8 (bottom panel). The CMEs are colored the same as in Figure 1 and the dashed colored lines represent the initial distance for each case. Kay et al. (2016) found that for V374 Peg, regardless of initial position, a single empirical function could describe the distance as a function of mass for all masses below $10^{17}$ g. Since the magnetic background of *κ¹ Ceti* is much weaker than that of V374 Peg we do not see such a clear trend. For epoch 2012.9, we see that the distance decreases significantly for the CMEs that begin the farthest from the ACS (purple, gray, and red). For the most massive of these cases we see that the final distance decreases with CME mass, but the trend breaks down for many of the lower-mass CMEs. These are the cases that begin rapidly deflecting toward the ACS and have enough momentum to continue propagating through the ACS. The other cases (dark and light blue, green, and orange) all have initial positions that are less than $10^0$ away from the ACS. For these cases we see small variations in the final position but no consistent trend. For epoch 2013.8 the initial distances tend to be slightly larger than those of epoch 2012.9. The final distance tends to decrease with CME mass until the least-massive cases, which show scattered behavior. The red cases have the smallest initial distance of the 2013 epoch cases and we see a slight increase in the final position of the most massive cases and a slight decrease in the final position of the least-massive cases, but all the deflections are no more than a few degrees.

We consider the average fractional change in the distance from the ACS, which we will use in our probabilistic model in Section 4. The average initial distance for all cases is $14.4^0$ for epoch 2012.9 and $21.9^0$ for epoch 2013.8. The corresponding average final distances are 7.4±4.4$^0$ and 10.3±9.3$^0$. We emphasize that the individual behavior varies significantly from CME to CME, but for both epochs, on average, the deflection tends to decrease the distance from the ACS by roughly a factor of two.

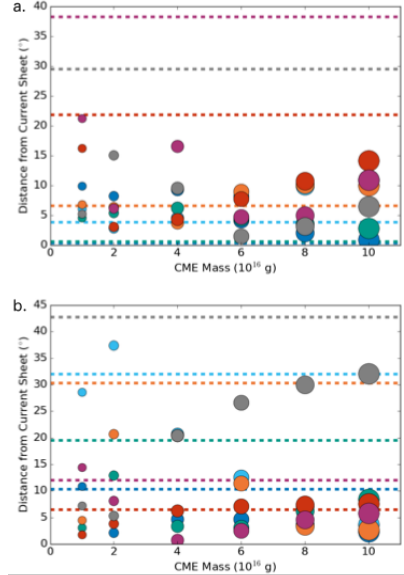

*Figure 2: Final distance from the ACS versus CME mass. The CMEs are colored the same as in Figure 1 and the dashed line represents the initial distance from the ACS for each color.*

### 4. The Probabilistic CME Impact Model

To better understand how CME deflection influences the rate of exoplanet impact we develop a simple probabilistic model Using Randomness to Simulate Impacts Near Exoplanets (*URSINE*). Given the number of CME eruptions per day we determine the probability of an eruption occurring in any time step. Using this probability and a random number generator we then determine when CMEs erupt over some time span. For each CME we randomly determine the properties beginning with a mass distribution and using the mass to scale the velocity and angular width.

We use a Gaussian distribution for log(mass) with a mean of 16.5 and a standard deviation of 1.15 as this represents a distribution in the range of masses we had previously considered (Kay et al. 2016). Vourlidas et al. (2010) fitted a Gaussian to the solar log(mass) distribution, which has different mean/standard deviation due to the less massive CMEs. We find that the frequency of impact is not particularly sensitive to small changes in the precise values used for the mean and standard deviation. We then randomly choose a position for the CME. If we have observations of the distribution of active regions we can incorporate it, but such is not the case for $\kappa^1$ *Ceti*. Instead, we assume that all longitudes are equally probable. We use 1 deg resolution for both the latitude range and the current sheet height (90x90 grid). For each latitude range/CS height we run 1000 CMEs (a total of 8.1 million CMEs in each panel).

Depending on the deflection model and CME shape the CME latitude and tilt may or may not be necessary. Here, we assume uniform probability of eruption within some latitude range and use a symmetric, cone-shaped CME so that the orientation is not needed. To describe the location of the ACS we use a sine wave with the amplitude reflecting the height from the magnetic field reconstruction. This amplitude will change over the stellar cycle. We then determine the CME's deflection, which is constrained by the *ForeCAT* simulations. This can be any simple model that depends on the CME properties and position and/or ACS position. Here we consider both a fixed final distance from the ACS and a model where the initial distance decreases by a factor of two. We then assume a planet orbiting at the equator and determine which CMEs produce impacts based on the relative planet and CME locations as a function of time.

All of the components of *URSINE* are easily replaceable with any other simple analytic or empirical models. Here, we use a uniform CME shape, similar to the cone model derived from fitting observations of solar CMEs (Xie et al. 2004). If the planet has an angular distance away from the CME nose less than or equal to the CME's angular width then we assume that a CME impact occurs. In a future work we will consider shapes that break the axisymmetry so that the rotation becomes important. The chosen deflection models do not utilize the CME mass but for other stellar systems there may be a clear dependence on the mass. In this work we focus simply on the fraction of CMEs that impact rather than their timing or even an absolute number of impacts. We do not need the CME speed, but this could be used to predict the arrival time of CMEs at the planet and could affect the precise timing of the impacts. The frequency of CME eruption from the star averages out when only considering the percentage of impacts so we simply consider enough CMEs to fully sample our distributions rather than trying to use the poorly constrained rate from stellar observations.

Based on the reconstructed magnetic field, we select ACS amplitudes of $25.8^0$ and $68.8^0$ for epochs 2012 and 2013, respectively. We first consider the deflection model where the CMEs are a fixed final distance from the ACS, using relatively close distances of $7.4^0$ and $10.3^0$ based on the *ForeCAT* models for epochs 2012.9 and 2013.8. From this we find a 29.6% chance of impact for epoch 2012.9 and a 16.2% chance of impact for epoch 2013.8, which we can understand through geometric arguments. The CMEs have angular widths between about $50^0$ and $70^0$, with the smaller values being more common. For the 2012.9 epoch the ACS is relatively flat so the CMEs all extend over the equator and the probability of impact is essentially the fraction of the total longitude extended by a CME's angular width at the equator. Given our angular width distribution, an equatorial CME covers roughly a third of all longitudes at the equator, which gives an upper limit of one-third for the probability of impact. In epoch 2013.8 the ACS is more inclined so the CMEs have larger absolute latitudes on average, decreasing their longitudinal extent at the equator, and causing the probability of impact to decrease by nearly a factor of 2.

For comparison, we can estimate the percentage chance of impact for current solar CMEs at Earth. Rather than comparing the total number of CMEs observed in the corona with the total number of near-Earth in situ observations we can use the frequency of "halo" CMEs. Halo CMEs have a projected angular width of $360^0$, which means they must be large and be propagating near the Sun-Earth line (when observed remotely by a near-Earth satellite) and are expected to hit Earth. Using observations from *Large Angle and Spectrometric Coronagraph Experiment (LASCO)* between 1996 and 2004, Lara et al. (2006) found that 3.5% of solar CMEs are halos but argued that, based on geometrical arguments, we should expect a value of 5.8%. We take 5% to be a reasonable order-of-magnitude estimate of the solar percentage and expect that the value varies over the solar cycle, but do not attempt to quantify that variation. For $\kappa^1$ *Ceti* we find that in epoch 2012.9 impacts are roughly five times as likely as solar-Earth impacts and three times as likely in epoch 2013.8.

We consider a second deflection model where the final CME distance from the ACS is half of its initial distance. This model requires assuming an initial latitude distribution of the CMEs - we assume a uniform probability over some latitude range and explore the sensitivity to the chosen value. For simplicity, we assume that the deflection only causes a change in the latitudinal direction so the longitude remains unchanged and the latitudinal distance is halved. We first use an initial latitude range of $70^0$, which causes impact probabilities of 29.3% for epoch 2012 and 25.5% for epoch 2013.8. Compared to the fixed ACS distance model the probability slightly decreases for epoch 2012.9 because a larger fraction of the final CME latitudes is at higher distances. For epoch 2013.8, some of the CMEs with final latitudes farther from the ACS are actually closer to the equator than in the fixed distance model due to the high inclination of the ACS, ultimately increasing the total probability of impact by a factor of 1.6 and making it more comparable to the 2012.9 epoch value. For comparison, we can determine the probability of impact without any deflections, which also depends on the initial latitude range but not the ACS inclination. For an initial latitude range of $70^0$ we find a 21.0% chance of impact so this deflection model causes the chance to increase by a factor of 1.4 for epoch 2012.9 (deflected percentage divided by non-deflected percentage) and 1.2 for epoch 2013.8. The percentage chance of impact depends on the initial latitude range. The closer the CMEs begin to the equator the more likely they are to impact a planet in an equatorial orbit. For epoch 2012.9, and using the halved-distance model, varying the initial latitude range between $30^0$ and $90^0$ causes the impact percentage to vary between 30.7% and 27.6%, a small but non-negligible effect. For epoch 2013.8 the impact percentage varies between 27.6% and 23.8%. The variations resulting from the change in initial latitudes are comparable to the effects of different ACS inclinations. The corresponding variations in the non-deflected percentages are from 30.2% to 16.3%. We can see that the change in the change is smallest for low inclination and low initial latitude range (factor of only 1.02) and largest for low inclination and high initial latitude range (factor of 1.7). For low initial latitude range and high ACS inclination we actually see a decrease in the chance of impact (factor of 0.9).

To further explore this relationship, we fully explore the parameter space of ACS inclination and initial latitude range. The resulting impact percentages are shown in Figure 3(a). Figure 3 confirms the results suggested by the previous test cases - both the ACS inclination and initial latitude range can cause changes of up to about 5% in the percentage chance of impact. The combined effect causes the impact percentages to vary between 21.6% and 31.3%. In Figure 3(b) we show the factor by which the percentage chance of impact increases or decreases when the effects of deflection are included. We take the percentage shown in Figure 3(a) and divide by the percentage chance without deflection, which only depends on the initial latitude range and not the ACS inclination. The contours are set so that white corresponds to no change, red is a decrease in the likelihood of impact, and blue is an increase in the likelihood of impact. As we saw before, we find that for low latitude ranges and ACS amplitudes the deflection has little effect on the chance of impact, high initial latitudes have an increase in the chance, and high ACS amplitudes decrease the chance. The ratio of the chances varies between 0.36 and 1.75.

The previous cases had an average angular width of $116^0$. We expect the results to be more sensitive to the deflection when the CMEs extend over a smaller fraction of three-dimensional space so we consider a second distribution with an average angular width of $60^0$. Figure 3(c) and (d) show the same as 3(a) and (b) but for the smaller angular width distribution. We see the same general behavior in the chance of impact with the values decreasing with both initial latitude range and ACS amplitude. The magnitude, however, is much smaller with the smaller angular width. The values in Figure 3(c) vary between 5.7% and 15.8%. The chance of impact falls off more rapidly with both parameters than in Figure 3(a). As expected, the effects of deflection are much stronger for the smaller CMEs with values varying between 0.36 and 2.02. Note that Figures 3(b) and (d) use the same color range for the contours. We see that there is strip where the initial latitude range and ACS inclination balance and cause the effects of deflection to be negligible, but the changes are much larger in the rest of parameter space. For a large ACS amplitude and low initial latitude range deflection pulls the CMEs away from the equatorial orbit of the planet, whereas with low ACS amplitude deflections compress the high initial CME latitude closer to the planet.

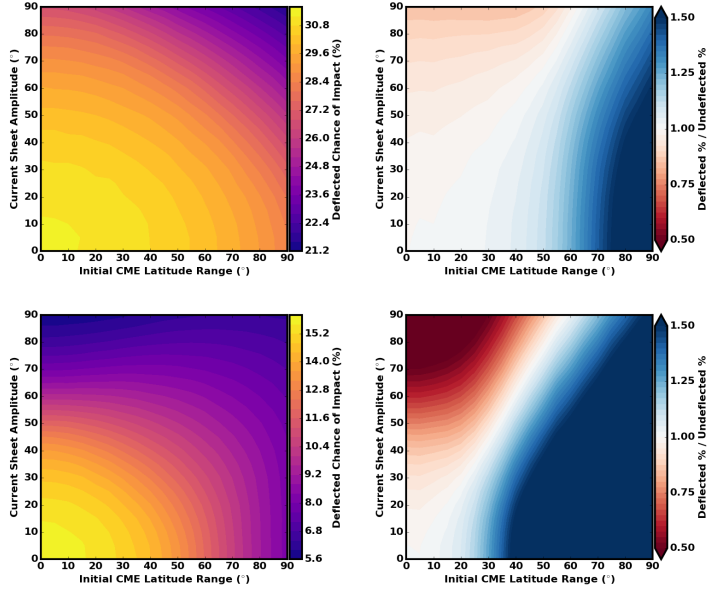

*Figure 3: The left panels show the percentage chance of impact as a function of ACS inclination and initial CME latitude range. The right panel shows the same, but normalized by the non-deflected values, highlighting the magnitude of the effects of deflection. The top panels show results for the standard angular width profile, and the bottom shows for smaller average angular widths.*

5. **Conclusion.**

Our *ForeCAT* model results suggest that the global magnetic background of the star is important in the determining the deflection and rotation of stellar CMEs, and thus, a critical factor in determination of CME collision frequency with planetary magnetospheres. We used the observationally constrained magnetic environment of $k^1$ *Ceti*, the young Sun's twin, at two epochs separated by 11 months and applied our 3D model, *ForeCAT* to study the trajectories of CMEs and derive the representative frequencies of CME collision with early Venus, Earth and Mars as well as exoplanets around young solar-like stars. Our results suggest that CME significantly deflect from the regions of strong magnetic field and coronal holes toward the ACS. Specifically, a magnetic configuration with slightly tilted dipole-like magnetic field favors the deflection of CMEs toward the ecliptic plane. Given that energetic CMEs have large angular width (with over $90^0$), this suggests that their distribution of their paths in interplanetary space should be highly anisotropic, and thus, can be considered to propagate in the ecliptic plane. Thus, the probability of the CME impact will increase up to 31.3%, which is a factor of 6 greater than that estimated by Airapetian et al. (2016). We should note that only compact CMEs are subject to deflection from the background field, while fully-blown CMEs do not experience deflection as they are the result of disruption of the entire streamer belt (Lynch et al. 2019).

A superflare with the energy of $2 \times 10^{34}$ erg have also been detected on the young Sun twin, $k^1$ *Ceti*, but the star has not been monitored extensively to derive the flare frequency (Schaefer et al. 2000). However, reliable estimates of the frequency of a Carrington-type flares (E ~ 2 x $10^{32}$ ergs) from a young, 300-700 Myr old solar-type stars like $k^1$ *Ceti*, come from the recent *Kepler*, *Gaia* and *TESS* data (Maehara et al. 2012; Notsu et al. 2019; Velloso et al. 2019). They suggest that young Suns generate flares with the maximum energy of ~ $10^{35}$ ergs. The frequency of Carrington-type flares (2 x $10^{32}$ erg) should be in the order of 200 events/yr or about 1 event/day. This suggests that the frequency of collisions of compact CMEs with the young Sun is ~ 0.3 events/day. We should note that fully-blown (halo) CMEs addressed by Lynch et al. (2019) are ejected in the region around ACS or at the frequency of one event per day. Because the time scale of a CME interaction with a planet is 2-3 days, these results imply that geoeffective Carrington-type CMEs were frequent events (one CME passage per day) in the history of early terrestrial planets and could have impacted their magnetospheres by opening the fraction of their magnetospheric filed by as much as 70%. These frequent events have played an important role in precipitation of high-fluence solar energetic particles into the lower atmospheres of early Venus, Earth, Mars and providing favorable conditions in production of biologically relevant molecules as well as forming nitrous oxide, the potent molecules for resolution of the long-standing Faint Young Sun (FYS) paradox for the early Earth and Mars (Fu et al. 2019; Airapetian et al. 2019a). Our results have direct implications for CME impacts with exoplanets around young and active G-K dwarfs and may suggest that Earth-sized exoplanets around late K and early M dwarfs could experience much more frequent interactions of CMEs with their magnetospheres, which may play a critical role in production of nitrous oxide, the potent greenhouse gas.

Finally, we want to emphasize the limitation of our model that is based on a PFSS model for the astrospheric magnetic field. If powerful large-scale compact CMEs are generated with high frequency, then the background magnetic field may not be transitioned to the equilibrium field and thus this approximation may not represent a realistic scenario. This would require time-dependent MHD simulations of CME interactions as it has been modeled in the solar corona (Lugaz et al. 2011,2012; Zuccarello et al. 2012, Lynch & Edmondson 2013, Zhou & Feng 2013, Zhou et al. 2014). Full MHD treatment of CME propagation in coronae of magnetically active stars will be performed in the near future.

**Acknowledgements**. V.S. Airapetian was supported by NASA Exobiology grant #80NSSC17K0463, Sellers Exoplanetary Environments Collaboration (SEEC) ISFM grant and *TESS* Cycle 1 project to study young solar analogs. The authors thank the referee for useful suggestions that improved this Letter.